# Spherical vs. Non-Spherical and Symmetry-Preserving vs. Symmetry-Breaking Densities of Open-Shell Atoms in Density Functional Theory

Shah Tanvir ur Rahman Chowdhury[1*],
and John P. Perdew[1, 2†]

[1]*Department of Physics, Temple University, Philadelphia, Pennsylvania 19122, USA,*
[2]*Department of Chemistry, Temple University, Philadelphia, Pennsylvania 19122, USA*

**Abstract**

The atomization energies of molecules from first-principles density functional approximations improve from the local spin-density approximation (LSDA) to the Perdew-Burke-Ernzerhof (PBE)) generalized gradient approximation (GGA) to the strongly constrained and appropriately normed (SCAN) meta-GGA, and their sensitivities to non-spherical components of the density increase in the same order. Thus, these functional advances increase density sensitivity and imitate the exact constrained search over correlated wavefunctions better than that over ensembles. The diatomic molecules studied here, singlet $C_2$ and $F_2$ plus triplet $B_2$ and $O_2$, have cylindrically symmetric densities. Because the densities of the corresponding atoms are non-spherical, the approximate Kohn-Sham potentials for the atoms have a lower symmetry than that of the external (nuclear) potential, so that the non-interacting wavefunctions are not eigenstates of the square of total orbital angular momentum, breaking a symmetry that yields a feature of the exact ground-state density. That spatial symmetry can be preserved by a non-self-consistent approach in which a self-consistent equilibrium-ensemble calculation is followed by integer re-occupation of the Kohn-Sham orbitals, as the first of several steps. The symmetry-preserving approach is different from symmetry restoration based upon projection. First-step space- (and space-spin-) symmetry preservation in atoms is shown to have a small effect on the atomization energies of molecules, quantifying earlier observations by Fertig and Kohn. Thus, the standard Kohn-Sham way of calculating atomization energies, with self-consistent symmetry breaking to minimize the energy, is justified, at least for the common cases where the molecules cannot break symmetry. Unless symmetry breaking is allowed in the molecule, PBE and SCAN strongly underestimate the atomization energy of strongly-correlated singlet $C_2$.

---

[*] Email: tanvir.chowdhury@temple.edu
[†] Email: perdew@temple.edu

1. INTRODUCTION

   Density-functional theory (DFT) is one of the most popular and successful quantum mechanical approaches to ground-state matter. It is nowadays routinely applied to calculate, e.g., the structures and binding energies of molecules in chemistry and of solids in physics. The atomization energies of molecules, or the energies needed to break all the bonds between the atoms, have long been important tests of approximate density functionals. These tests are straightforward when all the atoms are closed-shell like *He* or closed-subshell like *H*, and when all the molecules are similarly closed-shell or closed sub-shell. When this is not the case, the standard self-consistent Kohn-Sham calculations require further justification (some of it to be provided here) beyond the level of their numerical success. In an open-subshell atom, the approximated Kohn-Sham one-electron potential can be non-spherical, unlike the spherically-symmetric external potential, leading to a self-consistent electron density with spatial symmetry lower than that of the exact (non-spherical) density. That is the only symmetry breaking discussed in most of this article. While symmetries can break over long time intervals in reality, making symmetry breaking real or at least revealing, a single main-group atom is expected to be too small and too normally correlated to exhibit real symmetry breaking [1, 2]. By imposing a partial (first-step) space- or space-spin symmetry preservation, and thus showing that the space- and space-spin symmetry-breaking by approximate density functionals is energetically small for open-shell atoms, we provide more justification for the use of atomization energies of molecules (and solids), as standardly calculated, to test the approximate functionals. Here "small" is in comparison to the energy change from a spherical (ensemble) density to a non-spherical one, and also in comparison to the error reductions in the sequence Hartree-Fock, LSDA, PBE, and SCAN.

   In 1964 Hohenberg and Kohn [3] showed that there exists a universal non-relativistic density functional $F[n]$, independent of the external potential $v_{ext}(\boldsymbol{r})$ (e.g., the attraction of the electrons to the nuclei), such that minimization of the sum

$$F[n] + \int n(\boldsymbol{r})\, v_{ext}(\boldsymbol{r}) d^3r,$$

subject to the constraint

$$\int n(\boldsymbol{r})\, d^3r = N,$$

yields the ground-state energy and electron density of a quantum-mechanical $N$-electron system moving in this external potential. The Hohenberg-Kohn existence theorem has motivated the search for practical approximations to $F[n]$. Kohn and Sham [4] showed that a large part of $F[n]$ could be constructed from self-consistent one-electron wavefunctions or orbitals that are eigenstates of a self-consistent one-electron effective Hamiltonian, leaving only the density functional for the exchange-correlation energy to be approximated. The computational cost of a Kohn-Sham calculation is far less than that of a correlated-wavefunction calculation, especially for large $N$. Kohn-Sham spin-density functional theory [5] proved to be more accurate than Kohn-Sham total-density functional theory, because of the extra information that it provides to the approximate functionals. Here we will work within Kohn-Sham spin-density functional theory, with the further common assumption that there is no spin-dependence in the external potential or in the electron-electron interaction (which we take to be Coulombic).

The original proof of the Hohenberg-Kohn theorem was restricted to non-degenerate ground states, and the set of densities over which to minimize was restricted to the ground-state densities for the class of scalar external potentials $v_{ext}(r)$. These restrictions were removed in the Levy proof [6], which starts from the variational principle for the many-electron wavefunction, then performs the search over wavefunctions in two steps: First over all wavefunctions that yield a given density, and then over all densities that come from any wavefunction (i.e., over all reasonable densities [7]). The Levy proof can be extended [8] from a constrained search over wavefunctions to a constrained search over ensembles, which yields the same ground-state energy but different density functionals and an electron density with the full symmetry of the external potential. A remaining question for the exact Kohn-Sham theory (but not one we will much consider here) is whether the ground-state density of the real system can be replicated by the ground-state density of a system of non-interacting electrons in an effective scalar external potential $v_{eff}(\mathbf{r})$. The answer to this question of non-interacting *v*-representability is yes for the ensemble search, but not necessarily always yes for the wavefunction search [8]. It is possible that the Kohn-Sham wavefunction of lowest interacting energy is a low-lying excited state of a non-interacting system. The wavefunction of the Kohn-Sham auxiliary system of non-interacting electrons is a single Slater determinant or a linear combination of a few such determinants that are degenerate at the non-interacting level. For an interesting discussion of degeneracy, near-degeneracy, and symmetry in density functional theory, see Ref. [9]. Here we will consider only atoms and molecules in equilibrium, but there are also interesting effects in the binding energy curves of molecules [9, 10].

Janak *et. al.* [11] provided evidence that, within the local spin density approximation (LSDA) [4,5] for the exchange-correlation energy, non-spherical corrections to the energy are quite small in spin-polarized calculations for first-row atoms and can be accurately calculated by first-order perturbation theory for cases where the corrections are significant (transition-metal atoms and non-spin polarized treatments). Variational considerations lead one to expect that removing the constraint of spherical symmetry would lower the atomic energy. Fractional occupation numbers arise naturally in an ensemble picture. Kutzler and Painter [12] found much larger non-spherical corrections to the energy and strongly improved atomization energies for molecules from an early constraint-based generalized gradient approximation (GGA).

A local exchange-correlation energy density and potential at a point in space depends only on the electron spin densities at that point [3,4]. In the past few years, advances have been made in the development of the computationally-semilocal GGA [13-17] and meta-GGA [18]. A semi-local exchange-correlation energy density depends not only on the density at the point of interest but also on the gradient of the density at that point, and possibly on further information such as the non-interacting positive kinetic energy density there. This development of density-functional theory improves the predicted binding energies of *sp*-bonded molecules [19]. The beyond-LSDA functionals considered here are the Perdew-Burke-Ernzerhof (PBE) [17] GGA and the strongly constrained and appropriately normed (SCAN) meta-GGA) [18]. Like LSDA, those functionals are constructed by the satisfaction of exact constraints and are not fitted to the properties of any bonded systems.

The paper seeks to address the following question: Should any symmetry be imposed on the electron densities of open-shell atoms in DFT? In the next section, we report results from several calculations with approximate density functionals which show that the total energies of non-spherical atoms are systematically lower than those for spherical atoms, a result which leads to appreciably improved molecular binding energies. Next, we touch on the issue of the symmetry of the Hamiltonian and the ground-state density. Ref. [20] suggests that the Kohn-Sham noninteracting wave function need not display the symmetries of the interacting wave function. However, it must produce the correct spin densities, which are influenced by symmetry. While the ground state density has the full symmetry of the Hamiltonian in thermal-equilibrium ground ensembles and non-degenerate pure ground states, our work leads us to conclude that when there are degenerate pure ground states the best approximate functionals imitate the constrained search over pure states and not that over ensembles. The symmetry of the interacting ground-state wavefunction can be broken by the approximate Kohn-Sham non-interacting wavefunction, but the energetic consequences of that symmetry breaking in atoms as found here are too small to be important.

Our work extends the 1987 work of Kutzler and Painter [12] to the more modern PBE GGA [17] and to the SCAN meta-GGA [18]. Like Ref. [12], we focus on the atoms and homonuclear diatomic molecules of the atoms B, C, O, and F. Those atoms have open *p* subshells and non-spherical ground-state densities. The work of Ref. [12] was done at a time when fully self-consistent calculations (now the norm) were uncommon. Instead of performing fully self-consistent calculations, Ref. [12] used an equilibrium ensemble to make a Kohn-Sham effective potential with the same spatial symmetry as the external potential. For an atom, this approach yields a spherical density and a spherical Kohn-Sham effective potential. Ref. [12] then constructed a possibly less symmetrical density (e.g., a non-spherical density for an atom) by making integer occupations of the Kohn-Sham orbitals from that ensemble Kohn-Sham potential, with that potential having spherical symmetry for atoms and cylindrical symmetry for diatomic molecules. For many open-shell cases, this is not the self-consistent broken-symmetry solution that yields the lowest energy. We will start with their approach, calling it NS or non-spherical in section 2. Later we will recognize in it the first steps of a general approach to symmetry preservation in density functional theory, and compare it to the now-standard self-consistent approach that can break symmetry. We will first discuss the energetic effects of spatial symmetry breaking in the atoms, utilizing the computational approach of Kutzler and Painter, and later we will extend their approach to the energetic effects of spin symmetry breaking in the atoms.

## 2. CALCULATIONS USING SPHERICAL AND NON-SPHERICAL ATOMIC DENSITIES

In this section, we assess the impact of spherical and non-spherical atomic densities on calculated energies, following the Kutzler-Painter [12] approach described at the end of the preceding section. Anticipating a later discussion, we will refer to their approach as first-step non-self-consistent space symmetry-preserving (Sym-P1). As in their work, real Cartesian orbitals are chosen (and spin symmetry breaking in closed sub-shell molecules is not allowed.) Table I presents the spherical and non-spherical energies for several functionals. In all cases, we construct the state of maximum possible $z$-component of total spin, which for non-interacting pure states we take to be a single Slater determinant [21]. The electron configurations in the valence-shell integer-occupation or pure-state scheme are $p_z^1$ for boron (B), $p_x^1 p_y^1$ for carbon (C), $p_x^1 p_y^1 p_z^2$ for oxygen (O), and $p_x^2 p_y^2 p_z^1$ for fluorine (F). The corresponding atomic densities have cylindrical symmetry about the $z$-axis. Likewise, the electron configurations in the valence-shell fractional-occupation or equilibrium-ensemble scheme are $p_x^{1/3} p_y^{1/3} p_z^{1/3}$ for B, and $p_x^{2/3} p_y^{2/3} p_z^{2/3}$ for C, all with $z$-component of spin $S_\alpha$. For O and F, perhaps it is more revealing to present the fractional occupations divided into $\alpha$ and $\beta$ spin channels ($S_\alpha$ and $S_\beta$). For the oxygen atom,

$$S_\alpha : p_x^1 p_y^1 p_z^1$$
$$S_\beta : p_x^{1/3} p_y^{1/3} p_z^{1/3} .$$

Likewise, for the fluorine atom,

$$S_\alpha : p_x^1 p_y^1 p_z^1$$
$$S_\beta : p_x^{2/3} p_y^{2/3} p_z^{2/3}.$$

The corresponding atomic densities have spherical symmetry.

All DFT calculations for atoms and molecules were carried out in NWChem [22] using the unrestricted Kohn-Sham approach, allowing for a spin-dependent exchange-correlation potential. NWChem permits the use of fractional occupation numbers without additional coding. For a given spin multiplicity 2S+1, the $z$-component of total spin was set to S. The basis set was 6-311++G (3df,3pd), which converges valence-electron energy differences in Kohn-Sham DFT. For atoms, the spherical potential from the fractional-occupation configurations was used to generate the integer-occupied $p$ orbitals. In other words, the same $p$ orbitals are used in the spherical (equilibrium-ensemble) and non-spherical (symmetry-preserving, as discussed later) calculations, and only the occupations are changed. The numerical integration necessary for the evaluation of the exchange-correlation energy implemented in NWChem uses an Euler-MacLaurin scheme for the radial components (with a modified Mura-Knowles transformation) and a Lebedev scheme for the angular components. We use two levels of accuracy (the "xfine" and "huge" grids) for the numerical integration to get the total energy target accuracy of $1 \times 10^{-8}$ and $1 \times 10^{-10}$ Hartree. The biggest relative difference in atomization energies between these two target accuracies is only a quarter of a percent even for SCAN (see Appendix B).

The maximum number of iterations is set to 100. It should be noted that SCAN, PBE and our LSDA agree exactly for all uniform spin densities. Our LSDA uses the exact exchange energy per electron of a uniform electron gas and the PW92 [23] parametrization of the correlation energy per electron for uniform spin densities.

**TABLE I.** Effect of the removal of the spherical approximation on the atomic energies of B, C, O, and F with three nonempirical density functionals (SCAN, PBE, and LSDA).
Sym-P1 stands for non-spherical (first-step non-self-consistent symmetry-preserving) and Spherical for spherical densities. Energies are in Hartree, unless otherwise specified. 1 Hartree = 27.21 eV.

| Atom | SCAN | PBE | LSDA |
|---|---|---|---|
| **B (Spherical)** | -24.6216 | -24.6032 | -24.3504 |
| **B (Sym-P1)** | -24.6364 | -24.6085 | -24.3520 |
| **Difference (eV)** | 0.40 | 0.14 | 0.04 |
| **C (Spherical)** | -37.8181 | -37.7903 | -37.4650 |
| **C (Sym-P1)** | -37.8343 | -37.7939 | -37.4644 |
| **Difference (eV)** | 0.44 | 0.10 | 0.02 |
| **O (Spherical)** | -75.0355 | -74.9933 | -74.5173 |
| **O (Sym-P1)** | -75.0620 | -75.0041 | -74.5188 |
| **Difference (eV)** | 0.72 | 0.29 | 0.04 |
| **F (Spherical)** | -99.7047 | -99.6542 | -99.0998 |
| **F (Sym-P1)** | -99.7328 | -99.6613 | -99.0979 |
| **Difference (eV)** | 0.76 | 0.19 | 0.05 |

From Table I, it is apparent that PBE and especially SCAN energies are lowered significantly when evaluated with non-spherical densities. Density sensitivity increases from LSDA to PBE to SCAN, as might have been expected from the fact that the LSDA exchange-correlation energy density at a position in space depends only on the local spin densities, while in PBE it depends also on the local density gradients and in SCAN it depends further on the local non-interacting positive kinetic energy density. As can be seen in Table I, within the PBE approximation oxygen shows the largest non-spherical effect, with the total energy in the non-spherical treatment lying 0.29 eV lower than the result in the spherical approximation. For the advanced semi-local functional SCAN, the atomic energies of all four atoms are significantly lowered by including the non-spherical corrections (relaxing the spherical constraint). The largest difference is observed in the fluorine atom, where the SCAN functional gives an energy lowering of about 0.76 eV for the non-spherical atom compared with that in the spherical approximation, while the smallest effect occurs in the boron atom; 0.40 eV. The results for LSDA and GGA are in good agreement with those of Ref. [12]. Furthermore, using the same functionals as in Ref. [12], we reproduced similar energies (Table VII in Appendix A). The similarity between our results and those of Ref. [12] provides support for the correctness of our computations.
     Sym-P1

**TABLE II.** Binding energies of first-row dimers using spherical (equilibrium-ensemble) and non-spherical (first-step non-self-consistent space symmetry-preserving or Sym-P1) atomic densities. These densities of the atoms are paired with the naturally cylindrical densities of the dimers. Energies in electron volts (eV). The reference atomization energies are those experimentally observed for the ground state [12,24]. (See Table III for the electronic configurations of the dimers.) MAE is mean absolute error, widely used in density functional theory and permitting comparison with Table IV. (To find the atomization energies from self-consistent calculations with SCAN, PBE, and LSDA, subtract twice the small "Difference" in Table V from the entry here for "Sym-P1 atoms". The errors of the self-consistent atomization energies are displayed in Fig. 1.)

| | | Sym-P1 atoms | | | Spherical atoms | | |
|---|---|---|---|---|---|---|---|
| **Dimers** | **Reference** | **SCAN** | **PBE** | **LSDA** | **SCAN** | **PBE** | **LSDA** |
| $B_2$ | 3.01 | 3.18 | 3.40 | 3.88 | 3.98 | 3.69 | 3.97 |
| $C_2$ | 6.22 | 5.37 | 4.50 | 5.36 | 6.25 | 4.69 | 5.33 |
| $O_2$ | 5.12 | 5.76 | 6.39 | 7.69 | 7.20 | 6.98 | 7.78 |
| $F_2$ | 1.60 | 1.81 | 2.42 | 3.46 | 3.34 | 2.81 | 3.36 |
| **MAE** | | 0.47 | 1.05 | 1.54 | 1.21 | 1.32 | 1.57 |

Passing to the atomization energies of molecules, we see from Table II that inclusion of both nonlocal and non-spherical corrections (or relaxing the spherical constraint) gives closer agreement between theoretical and experimental binding energies of the first-row dimers. The binding energies are calculated from the experimental ground-state configurations of the molecules ($^3\Sigma_g^-$, $^1\Sigma_g^+$, $^3\Sigma_g^-$, $^1\Sigma_g^+$ for $B_2, C_2, O_2$ and $F_2$ respectively). Furthermore, the employed bond lengths of $B_2$, $C_2$, $O_2$ and $F_2$ are 1.59 Å, 1.243 Å, 1.208 Å, and 1.412 Å respectively [24]. Table II clearly illustrates that, while non-sphericity alone brings some improvement in calculated molecular binding energies, the use of nonlocal functionals in the atom calculations leads to significant further reductions in the errors. In fact, SCAN, when combined with a non-spherical density, produces the lowest mean absolute error or MAE (0.47 eV). In contrast, for a non-spherical density, the PBE functional yields an MAE of 1.05 eV and LSDA yields a large MAE of 1.54 eV. Furthermore, comparison between the non-spherical and spherical results for the same functional shows that non-spherical densities almost always result in an atomization energy closer to the experimental value. A possible inference from the results is that the sequence of approximate functionals LSDA, PBE, and SCAN is converging toward the exact density functional defined by a constrained search over wavefunctions, and not to the one defined by a constrained search over ensembles. One reason might be that the lower-symmetry densities of wavefunctions provide more information to the functional than do the higher-symmetry densities of some ensembles. A similar argument explains why approximate spin-density functionals are more accurate than approximate total-density functionals, even in systems where the external potential is spin-independent.

Figure 1 shows the errors of self-consistently calculated atomization energies, as given by the Sym-P1 atomization energies of Table II minus twice the small "Difference" from Table V. Note the improvement from LSDA to PBE to SCAN for the atomization energies in Fig. 1: While the change from LSDA to PBE is roughly a constant shift toward weaker bonding, the change from PBE to SCAN is more properly system specific.

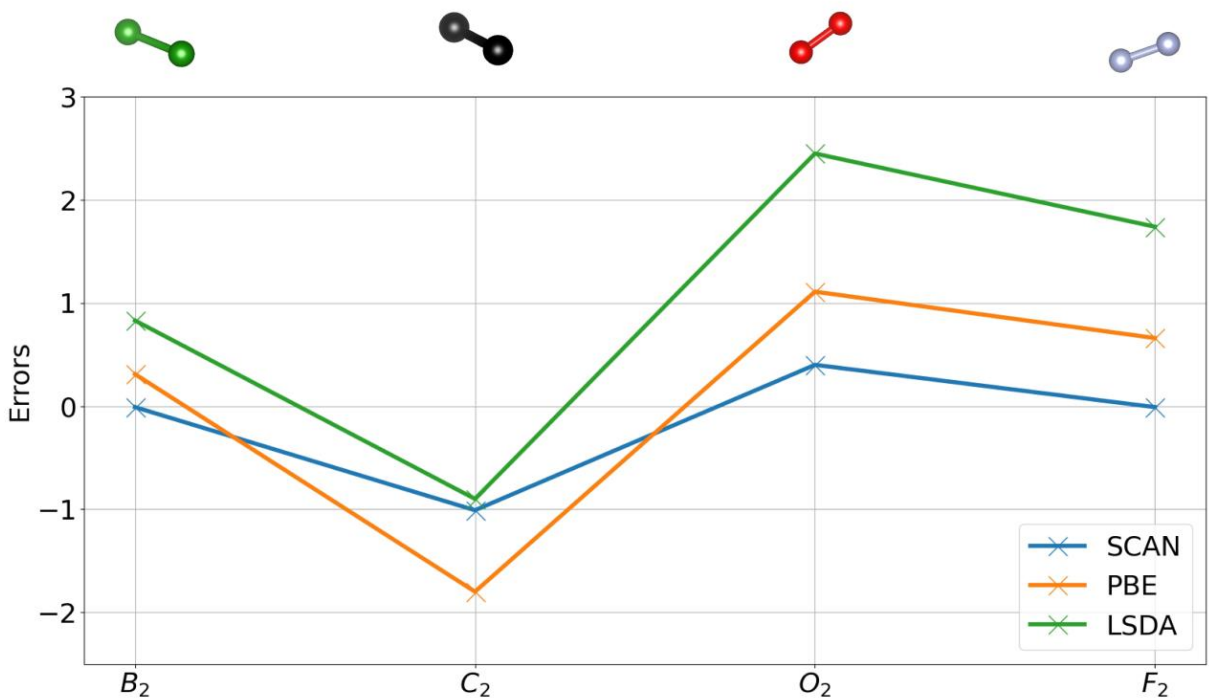

**FIGURE 1.** Errors (in electron volts) of the self-consistently calculated atomization energies of four diatomic molecules from the LSDA, PBE, and SCAN density functionals, in electron volts. The mean absolute errors (MAE) are 0.36 eV (SCAN), 0.97 eV (PBE), and 1.21 eV (LSDA). Spin symmetry breaking in strongly correlated $C_2$ has been suppressed, but if included it would reduce the errors of PBE and SCAN.

What is the effect of breaking the symmetry of the atoms and molecules considered here? The ground state electronic configurations of our molecules are illustrated in Table III. All the molecules studied here have $D_{\infty h}$ point symmetry group [25, 26] and have cylindrically symmetric ground-state densities. Thus, they have the symmetry of the external potential. For them, there is no difference in our calculations among the ensemble cylindrically symmetric, the symmetry-preserving, and the self-consistent densities.

**TABLE III.** Valence ground-state configurations [26] of $B_2, C_2, O_2$ and $F_2$ (dimers with space symmetry-preserving, cylindrically symmetric pure-state densities) along with their spin multiplicities.

|     | Valence Ground-State Configurations | Spin Multiplicity |
| --- | --- | --- |
| $B_2$ | $1\sigma_g^2 \ 1\sigma_u^2 \ 1\pi_{ux}^1 \ 1\pi_{uy}^1$ | 3 |
| $C_2$ | $1\sigma_g^2 \ 1\sigma_u^2 \ 1\pi_{ux}^2 \ 1\pi_{uy}^2,$ | 1 |
| $O_2$ | $1\sigma_g^2 \ 1\sigma_u^2 \ 2\sigma_g^2 \ 1\pi_{ux}^2 \ 1\pi_{uy}^2 1\pi_{gx}^1 \ 1\pi_{gy}^1$ | 3 |
| $F_2$ | $1\sigma_g^2 \ 1\sigma_u^2 \ 2\sigma_g^2 \ 1\pi_{ux}^2 \ 1\pi_{uy}^2 1\pi_{gx}^2 \ 1\pi_{gy}^2$ | 1 |

The case of $C_2$ is, however, intriguing. The straightforward application of the *aufbau* or building-up principle suggests that the ground-state configuration of $C_2$ at the equilibrium geometry is a singlet ($^1\Sigma_g^+$) configuration $1\sigma_g^2 \ 1\sigma_u^2 \ 1\pi_{ux}^2 \ 1\pi_{uy}^2$ [26], which is also cylindrically symmetric. However, we are dealing with a many-electron molecule, and the occupation of the lowest energy orbitals does not necessarily lead to the lowest energy. There is a possibility that excitation of an electron to a nearby orbital might lower the electron–electron repulsion and result in a lower overall energy despite the occupation of a higher energy orbital. The resulting configuration is a triplet ($^3\Pi_u$) with configuration $1\sigma_g^2 \ 1\sigma_u^2 \ 1\pi_{ux}^2 \ 1\pi_{uy}^1 \ 2\sigma_g^1$ [9, 26], which is not cylindrically symmetric (for real orbitals). Therefore, Ref. [9] suggests that singlet and triplet states compete in energy for $C_2$. This competition seems to be confirmed in the CCSD(T) total energies with the cc-pCVTZ basis set reported in Ref. [24]. But the story of singlet $C_2$ is complicated even further by its multi-reference strong correlation attributed [27] to an avoided crossing between two states of the same symmetry near the equilibrium bond length. Only some of this strong correlation can be captured by SCAN, and less by PBE, leading to an unusually strong underbinding of singlet $C_2$ by both functionals when the spin symmetry of the molecule is not allowed to break. In the full configuration interaction quantum Monte Carlo calculation of Ref. [27], the more normally correlated triplet state is a very low-lying excitation, lying vertically above the singlet by only about 0.2 eV.

Spatial symmetry does not break in our approximate Kohn-Sham descriptions of our cylindrical molecules. Spatial symmetry can break in the non-spherical open-subshell atoms B, C, O, and F, but as we will see the energy consequences of that are small.

# 3. WHAT DOES THE SYMMETRY OF THE HAMILTONIAN SAY ABOUT THE SYMMETRY OF THE GROUND-STATE DENSITY?

In this section, we summarize the known relationships between symmetry and degeneracy which are applied in this work. A symmetry operator $\hat{U}$ of a Hamiltonian operator $\hat{H}$ is a unitary operator that leaves the Hamiltonian invariant:

$$\hat{U}\hat{H}\hat{U}^\dagger = \hat{H}.$$

Since $\hat{U}$ is unitary, $\hat{U}^\dagger = \hat{U}^{-1}$ and $\hat{U} = e^{i\hat{A}}$ where the operator $\hat{A}$ is the self-adjoint generator of $\hat{U}$. Clearly, $\hat{U}$ and $\hat{A}$ must commute with $\hat{H}$. Now suppose that the Hamiltonian $\hat{H}$ of a system is invariant under a set of symmetry operators $\{\hat{U}\}$. Then any of these symmetry operators acting on a ground-state (GS) wavefunction $\psi_{ig}$ yields either the original GS wavefunction or another that is degenerate with it:

$$\hat{H}\psi_{ig} = E_g \psi_{ig}, \qquad \hat{U}\hat{H}\psi_{ig} = E_g \hat{U}\psi_{ig}, \qquad \hat{H}\hat{U}\psi_{ig} = E_g \hat{U}\psi_{ig}.$$

Let the set $\{\hat{U}\}$ of symmetry operators and their inverses be closed under multiplication, forming a symmetry group [26]. Any linear combination of degenerate eigenstates of $\hat{H}$ is another degenerate eigenstate, and the space of degenerate eigenstates is spanned by $N_g$ orthonormal degenerate ground states that form the basis for a symmetry-invariant subspace [28] of the state space, and the basis for an $N_g$-dimensional irreducible representation of the symmetry group [28]. $N_g$ is the degeneracy of $E_g$.

The statistical density operator for the ground state in the microcanonical (maximum entropy) ensemble at zero temperature is

$$\frac{1}{N_g}\sum_{i=1}^{N_g}|\psi_{ig}\rangle\langle\psi_{ig}|,$$

where $1/N_g$ is the probability of finding the system in the $i$th ground state. This is a product of a constant and the projection operator onto the subspace of degenerate ground-state wavefunctions, which is invariant when the same symmetry operator $\hat{U}$ is applied to all the ground-state wavefunctions. Starting from one ground-state wavefunction, all of those that are degenerate with it by symmetry (and not accidentally) can be generated by applying the symmetry operators to it. In this sense, "the symmetry of the ground state is the symmetry of the Hamiltonian" [29], and the symmetry of the ground-state density is also the symmetry of the Hamiltonian. An important special case occurs when the ground-state is non-degenerate, as for typical closed-shell systems Then, in a stronger sense (i.e., for individual ground-state wavefunctions), the symmetry of the ground-state density is the symmetry of the Hamiltonian. While atoms that form chemical bonds are typically open-subshell, and their pure states may not have the spherical symmetry of the atomic Hamiltonian, the molecules that they form are typically closed-subshell (but not in every case).

The symmetry operators in $\{\hat{U}\}$ may not all commute with each other, but those that do commute with each other can still be diagonalized along with $\hat{H}$. This is the typical case for open-shell non-relativistic (Coulomb interacting) atoms, where we can simultaneously diagonalize $\hat{H}, \hat{S}^2, \hat{S}_z, \hat{L}^2$, and $\hat{L}_z$. Since the energy of a real atom depends on the quantum numbers $S$ and $L$ that determine the eigenvalues $S(S+1)$ of the operator $\hat{S}^2$ and $L(L+1)$ of the operator $\hat{L}^2$ [29], the true ground-state wavefunctions must also be eigenstates of the operators $\hat{S}^2$ and $\hat{L}^2$. They can be but they do not have to be chosen as eigenstates of $\hat{S}_z$ and $\hat{L}_z$ for any choice of the z-axis. Thus the degeneracy of a multiplet is (2S+1)(2L+1). For a given electronic configuration at the non-interacting level, the ground multiplet should have maximum possible S, and for that S maximum possible L (Hund's rule). The symmetry operators for rotation through angle $\varphi$ about the z-axis in coordinate space are $e^{i\varphi \hat{L}_z}$, and in spin space $e^{i\varphi \hat{S}_z}$ [29]. The ground-state density will have the full symmetry of the Hamiltonian in thermal-equilibrium ground ensembles and non-degenerate pure ground states, but not necessarily for degenerate pure ground states.

Kohn-Sham ground-state spin-density functional theory in principle predicts the ground-state electron spin density and total energy of a real electronic system in the presence of a multiplicative and possibly spin-dependent external potential. The Kohn-Sham non-interacting wavefunction, often taken to be a single Slater determinant of spin orbitals that are eigenstates of the z component of an electron's spin, is intended to reproduce the spin densities (and non-interacting kinetic energy) of the interacting ground state, but it has been suggested that it need not otherwise be regarded as an approximation to a true ground-state wavefunction. From this perspective [20], it is hard to see why the Kohn-Sham wavefunction should be constructed as an eigenstate of the operators $\hat{S}^2$ and $\hat{L}^2$. While "spin contamination" can be removed from a wavefunction by projection (Ref. [30] for solids, and references therein), that approach seems better justified in wavefunction theory than in DFT. For an atom, the Kohn-Sham non-interacting wavefunction needs to be constructed as an eigenstate of the operators $\hat{S}^2$ and $\hat{L}^2$ only when that is required for the construction of an exact ground-state density. But that may in fact be required if we want a simple way to avoid symmetry-broken densities. It is known that singlet spin states of unbroken symmetry cannot have non-zero net spin densities, while symmetry-broken singlets have them. The work of Fertig and Kohn [31] suggests that eigenstates of $\hat{L}^2$ also have characteristic features in their densities.

Fertig and Kohn [31] argued that the total density of electrons in an atom with quantum number $L$ can be expanded as a sum of spherical harmonic contributions with even $l$ in the range $0 \leq l \leq 2L$. That would be true both for the exact electron density and for a density constructed in a non-self-consistent Kohn-Sham approach using a spherically averaged Kohn-Sham potential. Although a self-consistent Kohn-Sham calculation with an approximate functional could bring spherical harmonic contributions with $l > 2L$ into the electron density, along with a non-spherical Kohn-Sham potential, they argued that those inappropriate contributions to the density would be small. They also showed that the exact Kohn-Sham potential of an atom is non-spherical, and has an expansion that includes non-zero spherical harmonics to infinite order. That the symmetry of the exact Kohn-Sham potential need not be that of the external potential seems to be consistent with Ref. [32]. Nevertheless, the most direct way to preserve the symmetry of the density is to constrain the approximate Kohn-Sham potential to have the symmetry of the external potential, and that is the approach we will follow here.

The exact density functionals constructed from a Levy constrained search over many-electron wavefunctions will have the exact degeneracies of the exact quantum mechanical problem (even in the Kohn-Sham scheme if all the densities of the degenerate ground states are non-interacting v-representable), but approximate functionals will not. In many cases, the approximate functionals will predict more accurate total energies for the densities of the broken-symmetry states. For example, the computationally efficient semilocal functionals best describe moderate correlation like that in the slowly-varying electron gas on which they are largely based. They can describe strong correlation in a symmetry-unbroken wavefunction only by converting it to moderate correlation in a symmetry-broken wavefunction, and density functional predictions of net spin densities can be re-interpreted as predictions of total density and on-top pair density [33]. How should we define the energy of an open-shell atom for the calculation of atomization energies of molecules and solids from approximate functionals? The standard approach is to choose the broken-symmetry atomic state of the lowest approximate energy. We see nothing wrong with that, and it can lead to very accurate atomization energies when an accurate approximation like SCAN is used, as shown in Table IV with data from Ref. [36]. From the viewpoint of the density functional variational principle, this is the right thing to do, since it minimizes the approximated total energy functional via fully self-consistent calculations with possibly non-spherical Kohn-Sham effective potentials. (The predictiveness of SCAN for molecules and solids extends well beyond the atomization energies of Tables II and IV, but is not the main point of this article.)

**Table IV.** Mean absolute errors (MAEs) of the atomization energy for the six AE6 [34] *sp*-bonded molecules ($SiH_4$, $SiO$, $S_2$, $C_2H_4$, propyne, $C_2H_2O_2$ glyoxal, and $C_4H_8$ cyclobutane), in electron volts. The AE6 set was chosen [34] to be representative of the 109 atomization energies in Database/3 [35], which includes our $O_2$ and $F_2$ but not our $B_2$ and $C_2$. For the atoms and molecules, the self-consistent approximate Kohn-Sham wavefunction of lowest approximate energy is used, without imposing symmetries. Errors decrease from Hartree-Fock (numerical results from Ref. [34]) to DFT, and from the first to the third rungs of Jacob's ladder (numerical results from Ref. [36]) of approximations to the density functional for the exchange-correlation energy. These approximations are not fitted to any bonded system. A similar comparison for the formation energies of the 223 G3 molecules (including many molecules much larger than those in AE6) can be found in Ref. [18].

| | |
|---|---|
| Hartree-Fock Exchange | 6.3 |
| LSDA Exchange-Correlation | 3.3 |
| PBE GGA Exchange-Correlation | 0.6 |
| SCAN meta-GGA Exchange-Correlation | 0.1 (~1%) |

Symmetry breaking in density functional theory often brings positive benefits, including more accurate energies from approximate functionals and physical insight into strong correlations that are only implicit in the symmetry-unbroken ground-state wavefunction but freeze out in the DFT total or spin densities [1, 2, 33, 38]. But symmetry breaking emerges with growing system size [1, 2], and is not expected to be important in small atoms. In the next section, we will show that spatial symmetry breaking in the first-row atoms occurs in approximate DFT, but that it can be energetically unimportant, and that the symmetries of the exact density can be preserved (not just restored by projection [30]) if that is needed.

## 4. HOW ENERGETICALLY IMPORTANT IS SYMMETRY-BREAKING IN AN OPEN-SHELL ATOM?

The symmetry of the many-electron Hamiltonian is the symmetry of the external potential, and the symmetry of the Kohn-Sham one-electron effective Hamiltonian is the symmetry of the Kohn-Sham effective potential. These symmetries are the same when the Kohn-Sham effective potential is constructed from an appropriately chosen equilibrium ensemble.

Following a path laid out by Ref. [12], we can preserve the spherical symmetry of the external potential (and thus of the full interacting Hamiltonian) for an atom in the Kohn-Sham potential (and effective Hamiltonian) by doing a self-consistent equilibrium-ensemble Kohn-Sham calculation with fractional occupation numbers. But the spherical densities do not belong to degenerate ground-state wavefunctions. The symmetry of the external potential dictates, but is not necessarily the same as, the possible symmetry of the wavefunction. So, we recover a symmetry-preserved density by replacing the fractional occupation numbers in the equilibrium ensemble by integers, as done in Ref. [12] and described in the first paragraph of section 2. That is only a first step toward symmetry preservation, but it is what we implement in our numerical calculations. The full process of symmetry preservation is presented in the next paragraph.

For atomic ground states, non-interacting many-electron eigenstates of $\hat{S}^2$ and $\hat{L}^2$ are not typically single Slater determinants. If we want to construct a single Slater determinant to represent a Hund's rule ground state in an atom, we need to focus on the states with $\hat{S}^2$ and $\hat{L}^2$ eigenvalues $M_S = \pm S$ and $M_L = \pm L$, since these have single fully detailed electronic configurations. The spin orbitals then must be constructed as complex, current-carrying products of radial functions and spherical harmonics (complex linear combinations of the first-step symmetry-preserving orbitals). Since the densities of the spherical harmonic orbitals have cylindrical symmetry about the *z* axis, the total density also has this symmetry (as it has in first-step symmetry preservation). The presence of a non-zero current density implies that meta-GGAs like SCAN must be implemented in a way [39] that is not standard in most electronic structure codes. We defer these later steps to future work.

We start with a self-consistent calculation of the spherical, spin-polarized equilibrium ensemble density. For example, we occupy the carbon atom with $p_x^{2/3} p_y^{2/3} p_z^{2/3}$ electrons in $S_\alpha$ and no electrons in $S_\beta$. Then we find the self-consistent Kohn-Sham potential that this density produces, which is also spherical, like the external potential of an atom. We take the atomic orbitals for that potential, then occupy them with integer occupation numbers. Finally, we compute the corresponding energy for these orbital occupations, for comparison with the energy of the self-consistent broken-symmetry solution. This procedure has been carried out for the oxygen atom with $p_x^1 p_y^1 p_z^1$ electrons in $S_\alpha$ and $p_x^{1/3} p_y^{1/3} p_z^{1/3}$ electrons in $S_\beta$, the boron atom with $p_x^{1/3} p_y^{1/3} p_z^{1/3}$ electrons in $S_\alpha$ and no electrons in $S_\beta$, the fluorine atom with $p_x^1 p_y^1 p_z^1$ electrons in $S_\alpha$ and $p_x^{2/3} p_y^{2/3} p_z^{2/3}$ electrons in $S_\beta$. Table V compares the first-step symmetry preserved and -broken energies of the atoms under study.

**TABLE V.** Effect of first-step non-self-consistent space symmetry preservation on the atomic energies of B, C, O, and F with the use of three density functionals SCAN, PBE and LSDA.
All densities are non-spherical. Sym-Br stands for symmetry-broken or self-consistent and Sym-P1 for first-step non-self-consistent space symmetry- preserved computations.
Energies are in Hartree, unless otherwise specified.

| Atom | SCAN | PBE | LSDA |
|---|---|---|---|
| **B (Sym-P1)** | -24.6364 | -24.6085 | -24.3520 |
| **B (Sym-Br)** | -24.6393 | -24.6100 | -24.3528 |
| **Difference (eV)** | 0.08 | 0.04 | 0.02 |
| **C (Sym-P1)** | -37.8343 | -37.7939 | -37.4644 |
| **C (Sym-Br)** | -37.8371 | -37.7953 | -37.4653 |
| **Difference (eV)** | 0.08 | 0.04 | 0.02 |
| **O (Sym-P1)** | -75.0620 | -75.0041 | -74.5188 |
| **O (Sym-Br)** | -75.0663 | -75.0071 | -74.5210 |
| **Difference (eV)** | 0.12 | 0.08 | 0.06 |
| **F (Sym-P1)** | -99.7328 | -99.6613 | -99.0979 |
| **F (Sym-Br)** | -99.7371 | -99.6644 | -99.1003 |
| **Difference (eV)** | 0.11 | 0.08 | 0.06 |

The result that emerges from the data is that the energy difference between first-step symmetry-preserved and symmetry-broken densities is small. Table V highlights that spatial symmetry breaking lowers the energy of an atom, as expected, but only inconsequentially, usually much less than the errors of approximate DFT atomization energies. The overall smallness of the energy differences quantifies the conclusions of Fertig and Kohn [31]. Thus, the standard way of calculating atomization energies from Kohn-Sham theory, employing a single Slater determinant with self-consistent symmetry breaking to minimize the energy, is supported by this investigation. Table V also shows that the energy differences grow from LSDA to PBE to SCAN, in keeping with the increasing density-sensitivity of the functionals as already noted in the discussion of Table I.

Next, we consider symmetry breaking in both space and spin. To make the Kohn-Sham effective potential spin-independent, like the external potential, we must begin with a spin-unpolarized equilibrium ensemble. We occupy the carbon atom with $p_x^{1/3} p_y^{1/3} p_z^{1/3}$ electrons in $S_\alpha$ and $p_x^{1/3} p_y^{1/3} p_z^{1/3}$ electrons in $S_\beta$. Then we take the atomic orbitals for that potential and occupy them with integer occupation numbers for the correctly spin-polarized atom. Finally, we compute the corresponding energy for these orbital occupations, for comparison with the energy of the self-consistent broken-symmetry solution. This procedure has been carried out for the oxygen atom with $p_x^{2/3} p_y^{2/3} p_z^{2/3}$ electrons in both $S_\alpha$ and $S_\beta$, the boron atom with $p_x^{1/6} p_y^{1/6} p_z^{1/6}$ electrons in $S_\alpha$ and $S_\beta$, and the fluorine atom with $p_x^{5/6} p_y^{5/6} p_z^{5/6}$ electrons in $S_\alpha$ and $S_\beta$. Table VI compares the symmetry-preserved and -broken energies of the atoms under study.

**TABLE VI.** Effect of first-step non-self-consistent space-spin symmetry-preserved atomic energies of B, C, O, and F atoms with the use of three density functionals SCAN, PBE and LSDA. All densities are non-spherical. Sym-Br stands for symmetry-broken or self-consistent and Sym-P1' for first-step space-spin symmetry-preserved computations. Note that the Sym-Br or self-consistent total energies are the same as in Table V, but the Sym-P1' total energies are higher than the Sym-P energies because more symmetries are preserved in Sym-P'. Energies are in Hartree, unless otherwise specified.

| Atom | SCAN | PBE | LSDA |
|---|---|---|---|
| **B (Sym-P1')** | -24.6358 | -24.6083 | -24.3518 |
| **B (Sym-Br)** | -24.6393 | -24.6100 | -24.3528 |
| **Difference (eV)** | 0.10 | 0.05 | 0.03 |
| **C (Sym-P1')** | -37.8315 | -37.7927 | -37.4630 |
| **C (Sym-Br)** | -37.8371 | -37.7953 | -37.4653 |
| **Difference (eV)** | 0.15 | 0.07 | 0.06 |
| **O (Sym-P1')** | -75.0585 | -75.0016 | -74.5157 |
| **O (Sym-Br)** | -75.0663 | -75.0071 | -74.5210 |
| **Difference (eV)** | 0.21 | 0.15 | 0.14 |
| **F (Sym-P1')** | -99.7319 | -99.6605 | -99.0969 |
| **F (Sym-Br)** | -99.7371 | -99.6644 | -99.1003 |
| **Difference (eV)** | 0.14 | 0.11 | 0.09 |

The averaged difference between these two approaches in Table VI is 0.15 eV for SCAN, 0.09 eV for PBE, and 0.08 eV for LSDA. Relative to the energy of our self-consistent calculation, spin-symmetry preservation raises the energies of our four atoms and would also raise the energies of our two triplet molecules.

A standard way to preserve spin symmetry is the restricted open-shell formalism, as described for Hartree-Fock theory in Ref. [40]. It employs the same up- and down-spin orbitals for spin-paired electrons in an open-shell configuration, like the approach in our Table V. The approach of Table V restores spin symmetries in a simple non-self-consistent way, and also makes a first step toward restoring spatial symmetries.

## 5. CONCLUSIONS

Approximate non-empirical density functionals become not only more accurate for total energies and their differences but also more sensitive to the density as we go from LSDA to PBE to SCAN. In this sequence, the functionals better approximate the exact constrained search over correlated wavefunctions, and not the exact constrained search over ensembles. (For open systems of fluctuating electron number, which can be described only by ensembles, the large errors made by such functionals for total energies and their differences have long been known [37].) For accurate atomization energies of molecules from these functionals, and especially from SCAN, the densities of the open-shell atoms should not be sphericalized (ensemble-averaged over degenerate states). We have also found (Fig. 1) that the improvement in self-consistently-calculated atomization energy from LSDA to PBE is mainly an overall reduction of overbinding, while that from PBE to SCAN is more properly system specific for the normally-correlated molecules studied here.

The work of Fertig and Kohn [31] suggests that, to yield features of the exact ground-state density, an approximate Kohn-Sham non-interacting wavefunction of an atom can be constructed as an eigenstate of the square of the total angular momentum operator, just as the true or interacting wavefunction is. The spatial symmetry of the interacting ground-state wavefunction can be preserved from fractional to integer occupation numbers. Full preservation of spatial symmetry requires other steps described in the third paragraph of section 4 but not implemented here. The density change from first-step symmetry-preserved to self-consistent symmetry-broken is much smaller than the change from spherical to first-step symmetry preserved. Importantly, the former density change yields an energy change for LSDA, PBE, and SCAN that is small compared to the error reductions in the sequence Hartree-Fock, LSDA, PBE, and SCAN. This finding quantifies a conclusion of Fertig and Kohn [31]. These results have further strengthened our confidence that self-consistent symmetry breaking is the best way to calculate energies and energy differences. However, if we are interested in spatially symmetry-unbroken densities of atoms, then, as shown in this paper, they do not change the atomization energies significantly. Spin symmetries may also be preserved, if so desired, by starting from a spin-unpolarized equilibrium-ensemble, giving the Kohn-Sham potential all the symmetries of the external potential. Even fully symmetry-preserved densities are not claimed to be better than the self-consistent ones, except that they have the same symmetries as the exact densities (and their underlying Kohn-Sham non-interacting wavefunctions have the same symmetries as the true interacting wavefunctions). The "Differences" in our Tables I and V are not intended to be error estimates. What they suggest to us is that atomization energies that non-self-consistently preserve or self-consistently break the symmetries of the density are energetically close. Our conclusions about the non-empirical functionals LSDA, PBE, and SCAN are more firmly founded than our conclusions about symmetry preservation, where we have only taken a first step. However, that first step can also be the last step for B (one $p$-electron outside a closed subshell) and F (one $p$-hole in a closed shell), since for atoms in those electronic configurations only $L = 1$ is possible.

It was also found here that SCAN, which is usually accurate for atomization energies, underestimates that of strongly correlated singlet $C_2$ by about one electron volt. Spin symmetry breaking in the molecule might help, as it does for singlet $Cr_2$ [41], but was not found in our self-consistent calculation that started from a spin-unpolarized density. Starting the self-consistency cycle from a spin-polarized density is expected to improve the singlet $C_2$ atomization energy through spin symmetry breaking in SCAN, as it does [42] for the PW91 GGA (which is very similar [17] to PBE). Non-collinear spin-symmetry breaking [43] and spatial symmetry breaking are also possible. It seems possible that the combination of a highly constrained functional like SCAN with full symmetry breaking might yield reliably accurate energetics.

Future work on the current topic might investigate symmetry breaking by SCAN in strongly-correlated singlet $C_2$, complete the symmetry preservation work for *sp* atoms by implementing the third paragraph of section 4, and extend the symmetry-preservation work to atoms with *d* and *f* electrons (where there are also issues of non-interacting pure-state v-representability of the exact ground-state density [44]).

**ACKNOWLEDGEMENTS**

We are grateful for very helpful comments from two referees. We gratefully acknowledge Biswajit Santra at Schrödinger and the NWChem forum (including Edoardo Apra, Niri Govind and Eric J. Bylaska) for their valuable suggestions and discussions. The work of S.T.u.R.C. and J.P.P. was supported by the U.S. National Science Foundation under grant No, DMR-1939528 (CMMT - Division of Materials Research, with a contribution from CTMC - Division of Chemistry). This research includes calculations carried out on Temple University's HPC resources and thus was supported in part by the National Science Foundation through major research instrumentation grant number 1625061 and by the US Army Research Laboratory under contract number W911NF-16-2-0189.

**DATA AVAILABILITY**

The data that supports the findings of this study are available within the article (and its appendixes).

## APPENDIX A

In this appendix, we present a comparison of our results with those of Ref. [12] using the same three functionals GGA-VWN [45, 46], GGA-PZ [45, 47], and LSDA-VWN [46]. Calculations reported in the Table are carried out with aug-cc-pvqz basis set.

**TABLE VII.** Sym-P1 stands for non-spherical (first-step non-self-consistent space symmetry-preserving). Energies are in Hartree.

| Atom | GGA – VWN | | GGA – PZ | | LSDA – VWN | |
|---|---|---|---|---|---|---|
| | Our data | Ref. [12] | Our data | Ref. [12] | Our data | Ref. [12] |
| **B (Spherical)** | -24.680 | -24.681 | -24.678 | -24.679 | -24.353 | -24.353 |
| **B (Sym-P1)** | -24.688 | -24.687 | -24.686 | -24.685 | -24.355 | -24.354 |
| **C (Spherical)** | -37.890 | -37.891 | -37.885 | -37.887 | -37.468 | -37.469 |
| **C (Sym-P1)** | -37.896 | -37.896 | -37.891 | -37.891 | -37.468 | -37.468 |
| **O (Spherical)** | -75.143 | -75.146 | -75.137 | -75.140 | -74.522 | -74.523 |
| **O (Sym-P1)** | -75.158 | -75.159 | -75.151 | -75.152 | -74.526 | -74.526 |
| **F (Spherical)** | -99.831 | -99.838 | -99.826 | -99.832 | -99.106 | -99.111 |
| **F (Sym-P1)** | -99.843 | -99.847 | -99.837 | -99.841 | -99.106 | -99.109 |

## APPENDIX B

**TABLE VIII.** The percentage change of SCAN atomization energies between the two finest grids available in NWChem. Sym-P1 stands for non-spherical (first-step non-self-consistent space symmetry-preserving). The percentage changes are calculated as $\frac{x-y}{x} \times 100\%$, where, x is the SCAN energy using grid 'huge' and y is the SCAN energy using grid 'xfine'.

| | Sym-P1 | Spherical |
|---|---|---|
| $B_2$ | 0.25% | 0.22% |
| $C_2$ | -0.05% | -0.01% |
| $O_2$ | 0.08% | 0.09% |
| $F_2$ | 0.17% | 0.14% |

## REFERENCES

[1] P. W. Anderson, More is different, Science 177, 393 (1972).

[2] J. P. Perdew. A. Ruzsinszky, J. Sun, N. K. Nepal, and A.D. Kaplan, Interpretations of ground-state symmetry breaking and strong correlation in wavefunction and density functional theories, Proc. Nat. Acad. Sci. USA 118, e2 017850118 (2021).

[3] P. Hohenberg and W. Kohn, Inhomogeneous Electron Gas, Phys. Rev. B 136, 864 (1964).

[4] W. Kohn and L. J. Sham, Self-consistent equations including exchange and correlation effects, Phys. Rev., 140 (4A), A1133 (1965).

[5] U. von Barth and L. Hedin, A local exchange-correlation potential for the spin-polarized case. I, J. Phys. C: Solid State Phys. 5, 1629 (1972).

[6] M. Levy, Universal variational functionals of electron densities, $1^{st}$-order density matrices, and natural spin orbitals, and solution of the *v*-representability problem, Proc. Nat. Acad, Sci, USA 76, 6062 (1979).

[7] J. E. Harriman, Orthonormal orbitals for the representation of an arbitrary density, Phys. Rev. A24, 680 (1981).

[8] M. Levy, Electron densities in search of Hamiltonians, Phys. Rev. A 26, 1200 (1982).

[9] A. Savin, On degeneracy, near degeneracy and density functional theory, in Recent Developments in Modern Density Functional Theory, Edited by J. M. Seminario Elsevier, Amsterdam. (1996).

[10] J. P. Perdew, What do the Kohn-Sham orbitals mean? How do atoms dissociate?, in *Density Functional Methods in Physics*, edited by R. M. Dreizler and J. da Providencia, NATO ASI ATO ASI Series Series Series B: Physics, Vol. 123 (Plenum, NY, 1985).

[11] J. F. Janak and A. R. Williams, Method for calculating wave functions in a non-spherical potential, Phys. Rev. B 12, 6301 (1981).

[12] F. W. Kutzler and G. S. Painter, Energies of atoms with non-spherical charge densities calculated with nonlocal density-functional theory, Phys. Rev. Lett. 59 (12), 1285 (1987).

[13] D. C. Langreth and M. J. Mehl, Beyond the local-density approximation in calculations of ground-state electronic properties, Phys. Rev. B 28, 1809 (1983).

[14] J. P. Perdew and Y. Wang, Accurate and simple density functional for the electronic exchange energy: Generalized gradient approximation, Phys. Rev. B 33, 8880 (1986).

[15]   J. P. Perdew, Density-functional approximation for the correlation energy of the inhomogeneous electron gas, Phys. Rev. B 33, 8822 (1986).

[16]   C. D. Hu and D. C. Langreth, A spin dependent version of the Langreth-Mehl exchange-correlation functional, Phys. Scr. 32, 391 (1985).

[17]   J. P. Perdew, K. Burke, and M. Ernzerhof, Generalized gradient approximation made simple, Phys. Rev. Lett. 77, 3865 (1996).

[18]   J. Sun, A. Ruzsinszky, and J.P. Perdew, Strongly constrained and appropriately normed semilocal density functional, Phys. Rev. Lett. 115, 036402 (2015).

[19]   A. D. Becke, Density functional calculations of molecular bond energies, J. Chem. Phys. 84, 4524 (1986).

[20]   J. P. Perdew, A. Ruzsinszky, L. A. Constantin, J. Sun, and G. I. Csonka, Some fundamental issues in the ground-state density functional theory: A guide for the perplexed, J. Chem. Theory Comput. 5, 902 (2009).

[21]   O. Gunnarsson and B.I. Lundqvist, Exchange and correlation in atoms, molecules, and solids by the spin-density functional formalism, Phys. Rev. B 13, 4274 (1976).

[22]   E. Aprà, E. J. Bylaska, W. A. de Jong, N. Govind, K. Kowalski, T. P. Straatsma, M. Valiev, H. J. J. van Dam, Y. Alexeev, J. Anchell, V. Anisimov, F. W. Aquino, R. Atta-Fynn, J. Autschbach, N. P. Bauman, J. C. Becca, D. E. Bernholdt, K. Bhaskaran-Nair, S. Bogatko, P. Borowski, J. Boschen, J. Brabec, A. Bruner, E. Cauët, Y. Chen, G. N. Chuev, C. J. Cramer, J. Daily, M. J. O. Deegan, T. H. Dunning Jr., M. Dupuis, K. G. Dyall, G. I. Fann, S. A. Fischer, A. Fonari, H. Früchtl, L. Gagliardi, J. Garza, N. Gawande, S. Ghosh, K. Glaesemann, A. W. Götz, J. Hammond, V. Helms, E. D. Hermes, K. Hirao, S. Hirata, M. Jacquelin, L. Jensen, B. G. Johnson, H. Jónsson, R. A. Kendall, M. Klemm, R. Kobayashi, V. Konkov, S. Krishnamoorthy, M. Krishnan, Z. Lin, R. D. Lins, R. J. Littlefield, A. J. Logsdail, K. Lopata, W. Ma, A. V. Marenich, J. Martin del Campo, D. Mejia-Rodriguez, J. E. Moore, J. M. Mullin, T. Nakajima, D. R. Nascimento, J. A. Nichols, P. J. Nichols, J. Nieplocha, A. Otero-de-la-Roza, B. Palmer, A. Panyala, T. Pirojsirikul, B. Peng, R. Peverati, J. Pittner, L. Pollack, R. M. Richard, P. Sadayappan, G. C. Schatz, W. A. Shelton, D. W. Silverstein, D. M. A. Smith, T. A. Soares, D. Song, M. Swart, H. L. Taylor, G. S. Thomas, V. Tipparaju, D. G. Truhlar, K. Tsemekhman, T. Van Voorhis, Á. Vázquez-Mayagoitia, P. Verma, O. Villa, A. Vishnu, K. D. Vogiatzis, D. Wang, J. H. Weare, M. J. Williamson, T. L.

Windus, K. Woliński, A. T. Wong, Q. Wu, C. Yang, Q. Yu, M. Zacharias, Z. Zhang, Y. Zhao, and R. J. Harrison, "NWChem: Past, present, and future", J. Chem. Phys. 152, 184102 (2020).

[23] J. P. Perdew, and Y. Wang, Accurate and simple analytic representation of the electron-gas correlation energy, Phys. Rev. B, 45(23), 13244 (1992).

[24] NIST Computational Chemistry Comparison and Benchmark Database, NIST Standard Reference Database No. 101, Release 21, edited by R. D. Johnson, III; http://cccbdb.nist.gov/ (2020).

[25] In the Schoenflies system, the $D_{nh}$ group consists of the operations present in $D_n$ together with a horizontal reflection, in addition to whatever operations the presence of these operations implies. An important example is the $D_{\infty h}$ group, to which uniform cylinders and homonuclear diatomic molecules belong [26].

[26] P. W. Atkins, and R. S. Friedman. Molecular Quantum Mechanics. Oxford University Press (2011).

[27] G. H. Booth, D. Cleland, A. J. W. Thom, and A. Alavi, Breaking the carbon dimer: The challenges of multiple bond dissociation with full configuration interaction with quantum Monte Carlo methods, J. Chem. Phys. 135, 084104 (2011).

[28] M. Hamermesh. Group Theory and Its Application to Physical Problems. Dover Publications, New York (2012).

[29] R. Shankar, Principles of Quantum Mechanics, Springer Science & Business Media (2012).

[30] K. Tada, S. Yamanaka, T. Kawakami, Y. Kitagawa, M. Okumura, K. Yamaguchi, and S. Tanaka, Estimation of spin contamination errors in DFT/plane-wave calculations of solid materials using approximate spin projection scheme, Chem. Phys. Lett. 765, 138291 (2021).

[31] H. A. Fertig, and W. Kohn. Symmetry of the atomic electron density in Hartree, Hartree-Fock, and density-functional theories, Phys. Rev. A, 62(5), 052511 (2000).

[32] A. Görling, Symmetry in density-functional theory, Phys. Rev. A 47, 2783 (1993).

[33] J. P. Perdew, A. Savin, and K. Burke, Escaping the symmetry dilemma through a pair-density interpretation of spin-density functional theory, Phys. Rev. A 51, 4531 (1995).

[34] B. J. Lynch and D. G. Truhlar, Small representative benchmarks for thermochemical calculations, J. Phys. Chem. A 107, 8996 (2003).

[35] B. J. Lynch and D. G. Truhlar, Robust and affordable multi-coefficient methods for thermochemistry and thermochemical kinetics: the MCCM/3 suite and SAC/3, J. Phys. Chem. A 107, 3898 (2003).

[36] P. Bhattarai, B. Santra, K. Wagle, Y. Yamamoto, R. R. Zope, A. Ruzsinszky, K. A. Jackson, and J. P. Perdew, Exploring and enhancing the accuracy of interior-scaled Perdew-Zunger self-interaction correction, J. Chem. Phys. 154, 094105 (2021).

[37] J. P. Perdew, R. G. Parr, M. Levy, and J. L. Balduz, Density functional theory for fractional particle number: Derivative discontinuities of the energy, Phys. Rev. Lett. 49, 1691-1694 (1982).

[38] Y. Zhang, J. Furness, R. Zhang, Z. Wang, A. Zunger, and J. Sun, Symmetry-breaking polymorphic descriptions of correlated materials without interelectronic U, Phys. Rev. B 102, 045112 (2020).

[39] J. Tao and J. P. Perdew, Nonempirical construction of current-density functionals from conventional density-functional approximations, Phys. Rev. Lett. 95, 196403 (2005).

[40] A. Szabo and N.S. Ostlund, *Modern Quantum Chemistry: Introduction to Advanced Electronic Structure Theory* (Macmillan, NY, 1982).

[41] B. Delley, A. J. Freeman, and D. E. Ellis, Metal-metal bonding in Cr-Cr and Mo-Mo dimers: Another success of local spin density theory, Phys. Rev. Lett. 50, 488 (1983).

[42] J. P. Perdew, J. A. Chevary, S. H. Vosko, K. A. Jackson, M. R. Pederson, D. J. Singh, and C. Fiolhais, Atoms, molecules, solids, and surfaces: Applications of the generalized gradient approximation for exchange and correlation, Phys. Rev. B 46, 6671 (1992).

[43] S. Yamanaka, T. Ohsaku, D. Yamaki, and K. Yamaguchi, Generalized spin density functional study of radical reactions, Int, J. Quantum Chem. 91, 376 (2003).

[44] E. Kraisler, G. Makov, N. Argamon, and I. Kelson, Fractional occupation in Kohn-Sham density-functional theory and the treatment on non-pure-state v-representable densities. Phys. Rev. A 80, 032115 (2009).

[45] GGA-VWN is the GGA of Refs, [14, 15] with its local correlation in the parametrization by Vosko, Wilk, and Nusair [46]. GGA-PZ uses instead the parametrization of Perdew and Zunger [47].

[46] S. H. Vosko, L. Wilk, and M. Nusair. Accurate spin-dependent electron liquid correlation energies for local spin density calculations: a critical analysis, Can. J. Phys., 58(8), 1200-1211 (1980).

[47] J. P Perdew, and A. Zunger. Self-interaction correction to density-functional approximations for many-electron systems, Phys. Rev. B, 23(10), 5048 (1981).